\documentclass[12pt,a4paper]{article}     

\usepackage{graphicx}
\usepackage{bm}
\usepackage{amssymb}
\usepackage{subfigure}
\usepackage{siunitx}
\newcommand{\numm}[1]{\num[round-precision=2,round-mode=figures,
     scientific-notation=true]{#1}}
\newcommand{\be}{\begin{equation}}
\newcommand{\ee}{\end{equation}}

\begin{document}

\title{Surface roughness evolution in a solid-on-solid model 
of epitaxial growth
}

\author{Petar Petrov\thanks{E-mail: peynov@fmi.uni-sofia.bg} \\
Faculty of Mathematics and Informatics, Sofia University\\ 5 James Bourchier blvd., 1164 Sofia, Bulgaria\\
        \and
       Daniela Gogova \\
     Leibniz Institute for Crystal Growth\\ Max-Born-Str. 2, 12489 Berlin, Germany  
}

\date{\today}

\maketitle

\begin{abstract}
The paper presents results from kinetic Monte Carlo simulations of kinetic surface roughening using an important and experimentally relevant model of epitaxial growth -- the  solid-on-solid model with Arrhenius dynamics. A restriction on diffusing adatoms is included allowing
hopping down only at steps of height one monolayer in order to avoid possibly unrealistic events of jumping from arbitrarily high steps.   Simulation results and  precise analytic expressions representing the time evolution of surface roughness do not depend on the substrate size and clearly put forward the conclusion that for any basic set of parameters the model approaches in asymptotic limit the usual random deposition process with growth exponent $\beta=1/2$. At high  temperatures, it is preceded by a long transient regime characterized by a smooth surface covered with porous pillars and described by a power law with $\beta=3/4$.  
\end{abstract}

\bibliographystyle{spbasic}
\section{Introduction}
Surface morphology in epitaxial growth is a characteristics which often has a strong impact on important technological properties of the films.
Two of the main analytic tools for quantitative description of its temporal evolution are surface width and height-height correlation function. The surface width $w$ can be defined in different ways depending on the  measure used to estimate surface height fluctuations. The most common definition of $w$, the root-mean-square roughness (RMSR), is implemented  in many experimental measurement techniques -- Scanning  Tunneling Microscopy, Atomic Force Microscopy, Stylus Profilometry, etc.
\par\noindent
RMSR is  defined by the expression
\be
w(L,t)=  \langle \, {\overline{[h(t)-\bar{h}(t)]^2}\,}^{1/2}\, \rangle, 
\ee
where $h(t)$ is the surface height at a time $t$, the overbars denote spatial average, and the angular brackets stand for suitable ensemble average over simulation or experimental trials on sample  domains of size $L\times L$. Instead of {\em mean height} $\bar{h}(t)$, one can  equivalently use the term {\em surface coverage} $\theta$.
\par 
One of the most important concepts concerning the global surface width $w$ is the famous Family-Vicsek dynamic scaling principle \cite{fami1} which states that $w$ is represented by the formula
\be
w(L,t)=L^\alpha f(t/L^{\alpha/\beta}),
\label{scaling}
\ee
where the scaling function $f(x)$ satisfies  $f(x)\sim x^\beta$ for $x\ll 1$ and $f(x)$ is approaching a constant for $x\gg 1$. Eq. \ref{scaling} means that
$w(L,t)\sim t^\beta$ at any time $t$ smaller than a crossover time $t_c\sim L^{\alpha/\beta}$
and $w(L,t)\sim L^\alpha$ for $t\gg t_c$.  {\em The growth exponent} $\beta$ describes $w$ in the initial period of surface roughening  while {\em the roughness exponent} $\alpha$ characterizes the roughness of the saturated surface. Alternatively,  the scaling of the correlation function $G(\mathbf{r},t)=\langle\,\overline{[h(\mathbf{x}+\mathbf{r},t)-h(\mathbf{x},t)]^2} \,\rangle$ can be investigated, where  $h(\mathbf x,t)$ is the surface height at a point $\mathbf x$ and time $t$ and $\mathbf x,\mathbf r \in \mathbb{R}^d$ (in case of $d$-dimensional substrate). 
The fundamental importance  of the scaling principle lies in the fact that it implies similarity of surface morphologies obtained  at different length scales and equal other parameters.
\par
In recent decades, much theoretical and experimental research has been concentrated on scaling behavior of surface width and correlation function in non-equilibrium conditions, with the main objective to predict and classify the  roughness evolution in  epitaxial models.  Growth systems of different kinds  have been examined by   kinetic Monte Carlo (KMC) simulations  (see e.g.  \cite{sarm,krug3,smil2,bart1,anna,henk,andr,marm,kond}) and the corresponding continuum field approaches based on partial differential equations (the models of Mullins-Herring (MH) \cite{mull,herr}, Edwards-Wilkinson (EW) \cite{edwa},  Kardar-Parisi-Zhang (KPZ) \cite{kardar1}, Villain-Lai-Das Sarma (VLDS) \cite{vill1,lai}, etc.).  
 As a result of this research we now know that important from experimental point of view growth systems can be grouped  into different {\em universality classes} according to their scaling parameters in the regime of non-equilibrium surface roughening. 
\par
Despite the intensive work on the subject, the general behavior
of surface roughening which depends on large number of growth parameters is not yet completely clear even for commonly used epitaxial processes. The simplest settings of the problem for which an exact solution is known  correspond to {\em the random deposition model}  \cite{weeks1,evans1} with $\alpha=\infty$ and $\beta=1/2$. 
To the same class belong models with nonzero intralayer diffusion  but no interlayer transport (for example, due to large step-edge barriers).
\par
Much more multifarious is the behavior of the kinetic roughening in the  presence of interlayer transport. It is usually described by power laws $w(\theta)\sim \theta^\beta$ where $\beta >0$ or by  logarithmic curves ($\beta =0$) as in Family \cite{fami2} and EW models for 2-dimensional (2D) substrates  \cite{brune,mich,evans2}. Among the first studied systems are the homoepitaxial metal systems with body-centered and face-centered  cubic crystal geometries. Particularly important for surface smoothing appears to be the process of {\em downward funneling} (DF) -- the atoms deposited at step edges funnel down until reaching prescribed adsorption sites in lower layers. At $T = 0 $\,K, DF dynamics produces EW asymptotic behavior $w(\theta)\sim log(\theta)$ \cite{evans3}. 
\par
In general, all groups of theoretical models differ in their relaxation rules  which  in turn determine the corresponding universality class. While only several universality classes are commonly known for their applicability in experimental work, the total number of  different dynamic scaling scenarios seems to be infinite due to infinite number of possible (artificial) ways of choosing  the smoothing rules in the model. 
\par
An important property of scaling theory is  that the time axis can be divided into periods where the surface roughness $w$ is represented by different (possibly unstable) power laws $w\sim \theta^{\tilde \beta}$. The exponent $\tilde \beta$,  taking its values in the interval $[0,1]$,  becomes usually smaller in every following period (the two extreme values 0 and 1 correspond to { the asymptotic steady state} and  {the steepest roughening}, respectively).
Universality classes are well-known with the set of exponents defining the final asymptotic steady state ($\theta \rightarrow \infty$) but any pre-asymptotic {\em transient}  period can also be very important for applications especially if it complies with experimental requirements. For example, every growth process with a fast enough surface relaxation usually begins with a 2D layer-by-layer mode that under certain conditions could be sufficient for the concrete experiment.
\par
In the early atomistic SOS models the surface is relaxed in such a way that the global roughness increases monotonically until reaches an asymptotic steady state determined by the substrate size. This is caused by the fact that after adsorption, the adatoms perform a single hop and move either directly to a site of local minimum height (EW-universality class) or to some of its nearest neighbors of any height  (VLDS-universality class).  Traditionally, the VLDS-universality class is considered as a correct enough description of conserved epitaxial growth processes, but there exist microscopic situations in which it seems to be unsatisfactory. A natural complication of early models is a KMC model based on Arrhenius activated diffusion, called in the literature also  {\em realistic} or {\em collective diffusion model} stressing on the fact that detailed motion of the deposited atoms is included. As it is pointed out in  \cite{kardar2}, the main difficulty with the analysis of RMSR in  realistic models is that all involved processes as diffusion, nucleation, atom incorporation at steps, etc., take place over a wide range of time scales. Such schemes are appropriate for simulations below the roughness transition allowing to avoid equilibrium surface roughening and ignore evaporation. A crucial parameter is the substrate size which not only determines the asymptotic state, but if it is not large enough compared to the adatom diffusion length,  can also be a limiting factor for obtaining the correct scaling behavior. 
\par
Several SOS models with Arrhenius dynamics are used in the literature for understanding the dynamic scaling of conserved epitaxial growth. 
One of the first growth models with detailed description of surface diffusion in 1+1D is published in \cite{tamb} where the effective exponents $(\alpha_{\hbox{\scriptsize eff}},\beta_{\hbox{\scriptsize eff}})$ are found to vary from $(\infty,1/2)$ (random deposition) to approximately $(0,0)$ with increasing the temperature, and at intermediate temperatures a possible growth exponent $\beta$ of a new universality class is proposed. The fundamental question for existence of a universality class which includes the SOS model with Arrhenius dynamics is addressed  in a number of works (see e.g. \cite{vill1,lai,krug2,smil2,kort,wil}). Most of the papers show temperature dependence of the effective growth exponent, crossovers from linear to nonlinear growth exponents and do not exclude the possibility for unstable growth. The membership of SOS model in  VLDS class is finally proved in   \cite{has1,has2,has3} and only recently, an analytic formula for $w$ is derived  in  \cite{aar} in the case of irreversible growth:
\be
w=\frac{L^\alpha}{R^{1/2}}f(Rt/L^{\alpha/\beta}),
\ee
with $\alpha=2/3$, $\beta=1/5$  (2+1D case, see also \cite{vill1,lai}).
\par
The present work is devoted to  evolution of surface roughness 
$w$  obtained by KMC simulations with an alternative to the commonly used SOS model presumably improving some of its potential "weaknesses"  and describing well enough  the conserved epitaxial growth with Arrhenius dynamics. 
\section{Motivation and a short description of the model}
In most SOS models with Arrhenius dynamics any adatom is allowed to perform a  hop down no matter how large the step height is and this  rule is the main factor for formation of  asymptotic steady state. In order to satisfy the requirement to produce no overhangs, the adatom should "slide" in one move only along a whole column side while eventually sits on top of the neighboring column. As it is noted in \cite{kort}, this is unrealistic and causes anisotropy between horizontal and vertical surface diffusion. Such a compromise with jumps down of arbitrary height has been used in simulations since models including detailed vertical diffusion would be much more time consuming. There exist also models keeping the height difference between any two adjacent columns smaller than a prescribed global number \cite{kim1,kim2}. Such a strategy avoids the  potential weakness of the model described above but apparently introduces another one and thus, also fails to identify the real class corresponding to SOS schemes with Arrhenius dynamics.
\par
Looking for a possibly better solution of the discussed problem we propose here the following rule for surface diffusion. Let us assume an atom is sitting at a step of height larger than 1 and is allowed to perform only moves down of height 1. If after the first vertical move down along the column side the atom possesses only lateral neighbors (at least two), this will lead with a high probability to an overhang which violates the condition for conservativeness. If it has only one lateral neighbor, the activation energy  for such a hop will be higher than activation energy in the case of one lateral and one vertical neighbor beneath because of the lower number of nearest neighbors at the final position of the move. If the difference between the two activation energies is high enough one can allow  interlayer transport  downwards only at steps of height 1. This is what is assumed in our model  (diffusion upwards is not allowed  and only  irreversible growth is investigated in the current settings).
\par
 Our simulations with relatively low values of diffusion-to-deposition ratio $R$ produce a  surface morphology characterized with formation of  voids, cracks, pits, etc. which is  known  often to occur in real growth conditions at low temperatures. 
With increasing of $R$, however, in long time periods the concentration of such irregularities is limited to a low value, eventually vanishing at $R\rightarrow \infty $ (see also Section~\ref{sec4}). 
\par
Generally, two main factors  are responsible for time evolution of surface morphology -- deposition competes with surface  diffusion. 
Since these mechanisms are controlled by different parameters, for a precise comparison with  experimental results, simulation and theoretical models  need to take into account at least a basic set of growth parameters which include  temperature $T$,  deposition rate $F$,  diffusion rate $D$ and coverage $\theta$. Following this strategy, we build a model based on a simple cubic geometry with the typical assumptions: the atoms adsorb  on top of  already deposited ones (thus, no overhangs are allowed) at a constant rate $F$; they diffuse on terraces randomly to one of its nearest neighbors at a rate $D$ (Eq.~\ref{diff}) and desorption has negligible values; for the sake of simplicity, in this  study only irreversible aggregation is considered -- the atoms become immobile after attaching to an island or another atom (corresponding to zero or negligible detachment rate); {\em an important difference with the previous conserved SOS models} is that the atoms can  move to another layer only by hopping {\em one monolayer (ML) down} with the same rate (0 for steps of height 1 and infinite value for higher steps of Ehrlich-Schw\"obel (ES) barrier); diffusion rate $D$ satisfies an  Arrhenius law:
\be
D = \nu \exp( -\Delta E/k_BT),
\label{diff}
\ee
where $\Delta E$ is the diffusion barrier for an adatom on a terrace.  The attempt frequency is chosen to be $\nu=2k_BT/h$ ($k_B$ is Boltzmann's constant, $h$ -- Planck's constant).
\par
Note that the models with a { single} dominating kinetic process differing in some of the parameters (temperature, activation energy, attempt frequency or deposition rate) can be unified by considering only the diffusion-to-deposition ratio (variable) $R=D/F$. Therefore, the surface evolution can be fully described by two independent variables - the coverage $\theta$ and $R$. 
The two extreme values of $R$ are $R\rightarrow 0$ ($\Delta E$, $F$ are sufficiently large or $T$ --  sufficiently low) and $R\rightarrow\infty$ ($\Delta E$, $F$ are sufficiently small or  $T$ --  sufficiently high). The latter case corresponds to an ideal smoothly growing film with an oscillating $w$ while the former one is { the random deposition model} (the "benchmark" model as it is called in \cite{evans2}, see also \cite{reif,weeks1,evans1}) in which the atoms are randomly deposited at a constant rate on top of the previously adsorbed ones, the columns grow independently and obey Poisson statistics:\\
The probability  $P_k(t)$ -- a given column to be of height $k$ at a time $t$  is
\be
P_k(t)=\frac{(Ft)^k}{k!}\exp(-Ft), 
\ee
the coverage $\theta=\bar{h}(t)$  is
\be
\bar{h}(t)=\sum_{k=0}^\infty k\frac{(Ft)^k}{k!}\exp(-Ft)=Ft,
\ee
and the RMRS $w$ is given by
\be
w(t)=\left (\sum_{k=0}^\infty \frac{(Ft)^k}{k!}\exp(-Ft)(k-Ft)^2\right )^{1/2}= (Ft)^{1/2}=\theta ^{1/2}.
\ee
Similar problems are considered in \cite{evans1} for different adsorption-site geometries and different structures. For adsorption at four-fold hollow sites of a face-centered or body-centered cubic geometry, the columns do not grow independently, and the asymptotic kinetic roughening has the form $w\sim \theta ^\beta$ where $\beta \approx 1/4$. 
\par
All simulations in our study are performed on a substrate of size  $100\times 100$, with deposition rate $F=0.01$\,ML/s, temperature $T=800\,^{\circ}\mathrm{C}$, $\Delta E$ mainly in the range $1.85-2.75$\,eV  and  time $t$ up to  $\numm{4.0e+06}$\,s. Translated to $(R,\theta)$-"terminology", this means $R=\numm{E+02}-\numm{E+07}$ and $\theta=0-\numm{4.0e+04}$.
The conclusions from our simulations are not hindered by the relatively moderate substrate size and coverage since the results presented below as well as single trials on larger substrates ($200\times 200$ and $300\times 300$)
  show no dependence of  RMSR on the system size which in turn means that the problem does not exhibit finite size effects. We measure the roughness at coverages $\theta$ with uniformly distributed fractional part and therefore, in the case of ideal layer-by-layer growth ($R\rightarrow\infty$), it is expected to oscillate between 0 and 1/2  with a mean value 
\be
\label{rough-infty}
w_\infty=\int_0^1\sqrt {x-x^2}dx\approx \numm{3.93e-1}. 
\ee
\section{Results and discussion}
\begin{figure}[ht!]
\begin{centering}
\includegraphics[width=0.49\hsize]{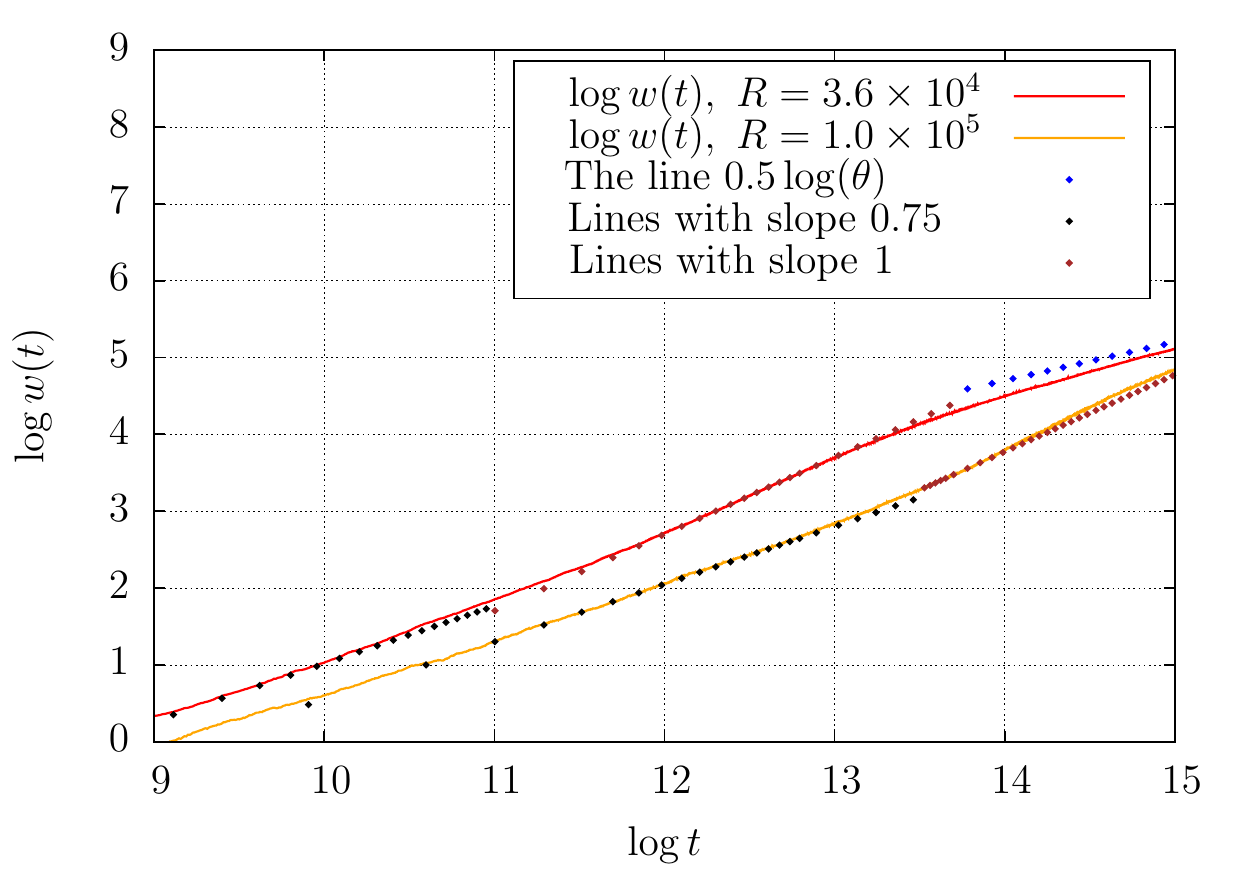} \hfill
\includegraphics[width=0.49\hsize]{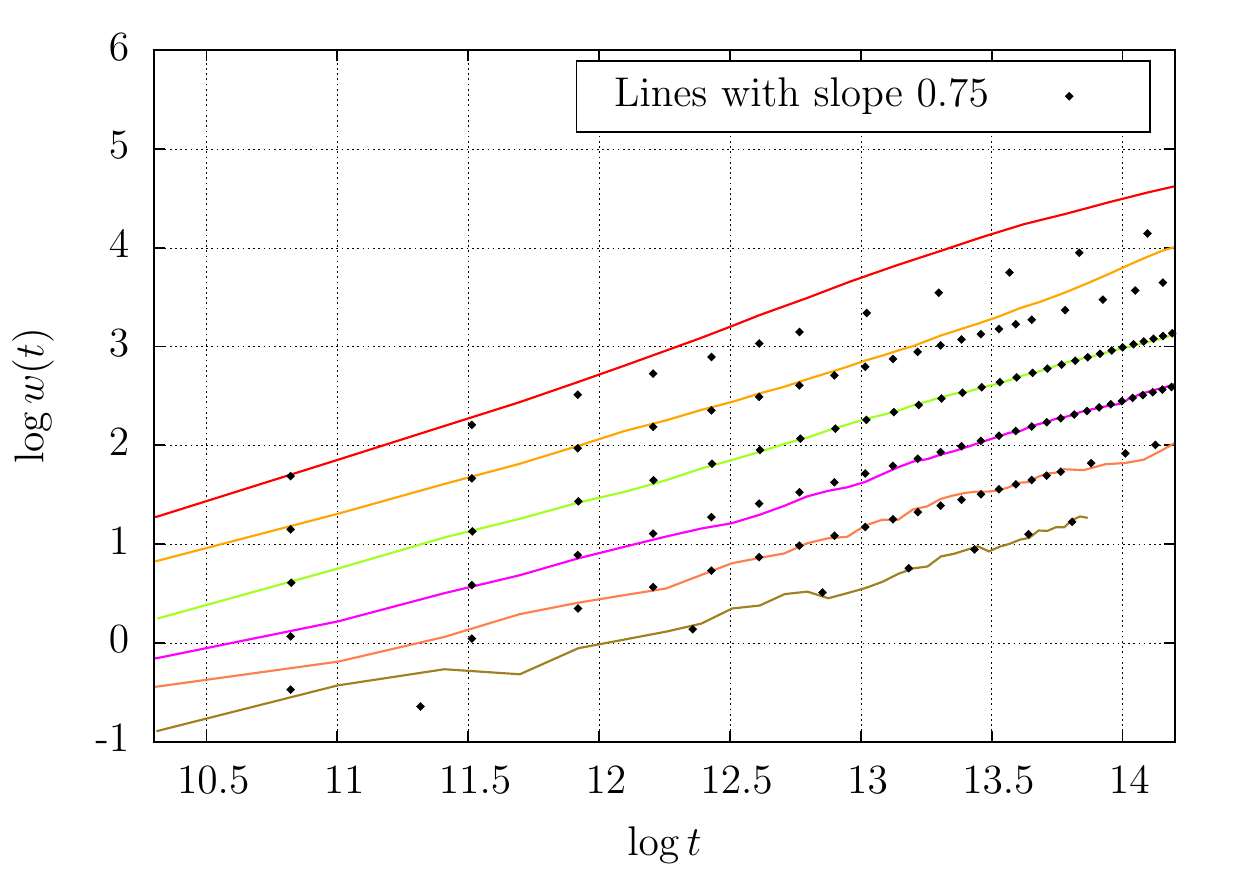}
\caption{\label{fig-feel}{Logarithmic profile of $w$ compared with gradients 0.5, 0.75 and 1 (left) and 0.75 (right). The left graph represents $w$ from simulations with $R=\numm{3.6E+04}$ and $R=\numm{1.0E+05}$ and the right -- with $R=\numm{3.6E+04}-\numm{7.9E+06}$.}}
\end{centering}
\end{figure}

In order to get a feeling of the effective growth exponent we first examine the curve of $\log w$ (shown in Fig.~\ref{fig-feel} (left) for two particular energies corresponding to $R=\numm{3.6E+04}$ and $R=\numm{1.0E+05}$). 
One can distinguish between three different growth modes.  After the initial increase of the growth exponent $\beta$ (not shown here), a stable period with $\beta =0.75$ follows which length is an increasing function of $R$ (see black dotted line). This is the first growth regime  (it can be seen also in Fig.~\ref{fig-feel} (right) for higher values of $R$). The second straight line with a slope 1 (brown dotted line) shows the existence of a second (short unstable) growth mode, where $\beta$ increases from 0.75 to values equal to 1 (for $R=\numm{3.6E+04}$) or greater than 1 (for $R=\numm{1.0E+05}$). The highest value of $\beta$ clearly corresponds to the inflection point of the function $\log w$ which changes from convex in the first period with $\beta =0.75$ to concave in a second period where the curve approaches the line $0.5\log \theta$. This suggest $\beta =0.5$ as a strong candidate for a growth exponent in the asymptotic state or equivalently, $w$ tends to the random deposition power function $\theta ^{1/2}$ as $\theta \rightarrow \infty$ and {\em in asymptotic sense, the model coincides with random deposition model}. 	Stronger evidences and precise expressions for $w$ as a function of $R$ and $\theta$ follow below.
\par
\begin{figure}[ht!]
\begin{centering}
\includegraphics[width=0.49\hsize]{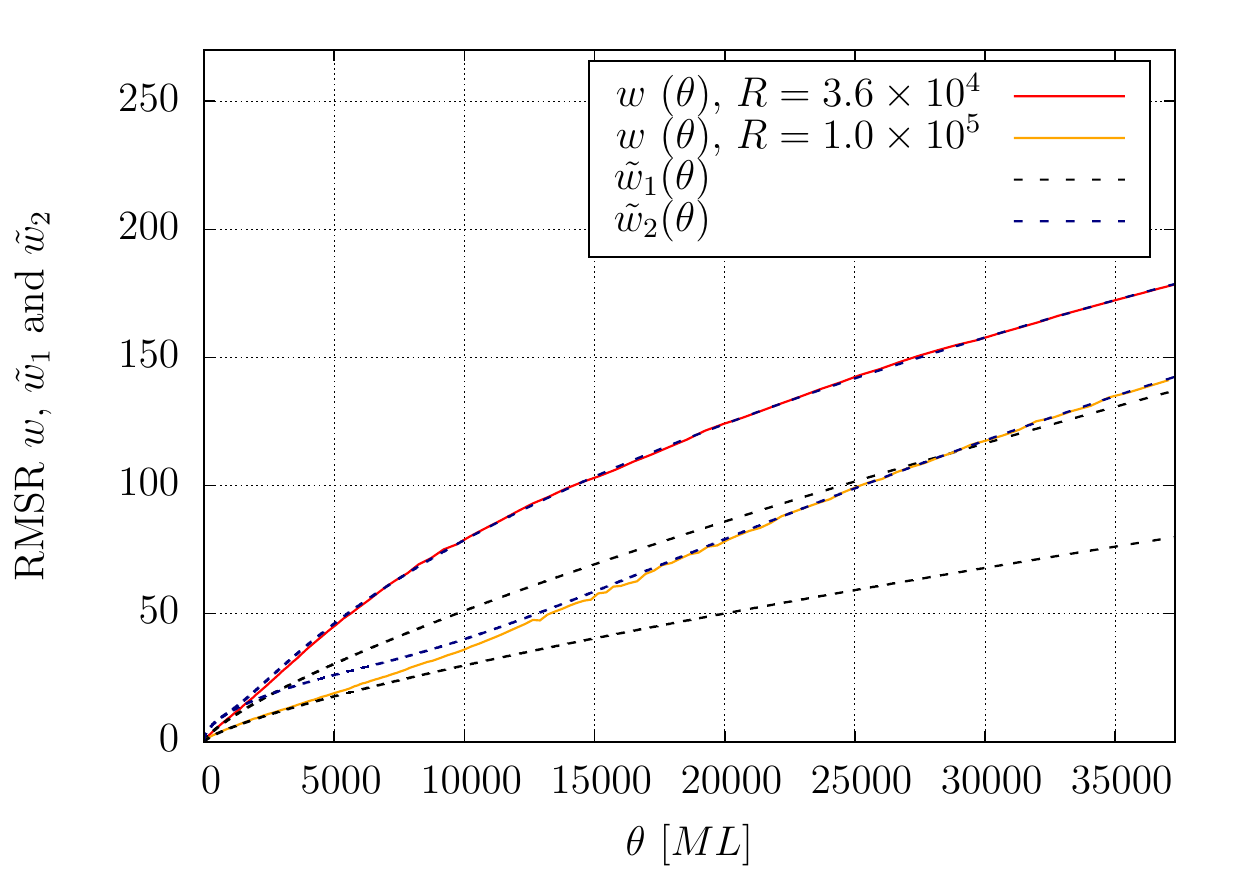} \hfill
\includegraphics[width=0.49\hsize]{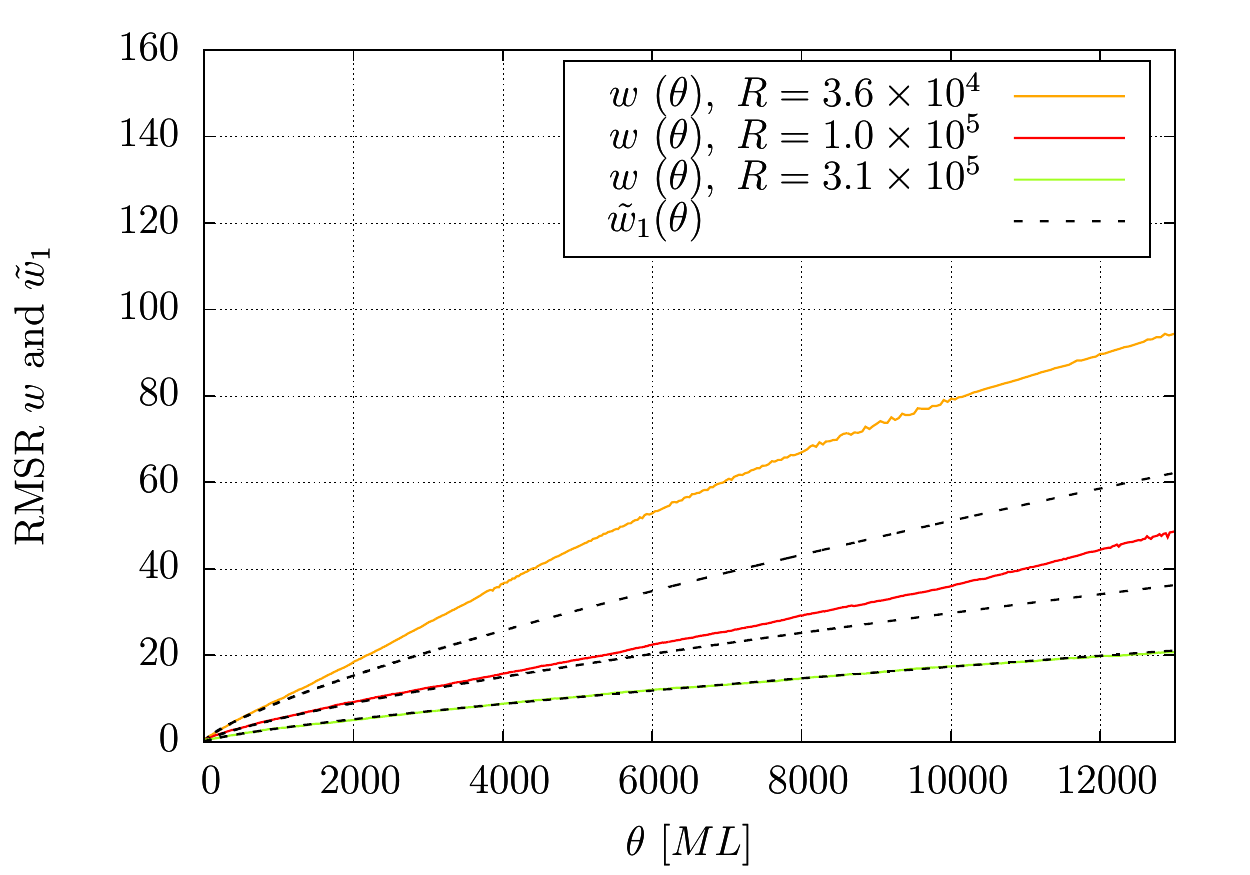} 
\caption{\label{fig-tot}{Approximation of $w$ simultaneously by $\tilde{w}_1$ and $\tilde{w}_2$  (a) and by  $\tilde{w}_1$  (b) for different values of $R$.}}
\end{centering}
\end{figure}

\begin{figure}[ht!]
\begin{centering}
\includegraphics[width=0.49\hsize]{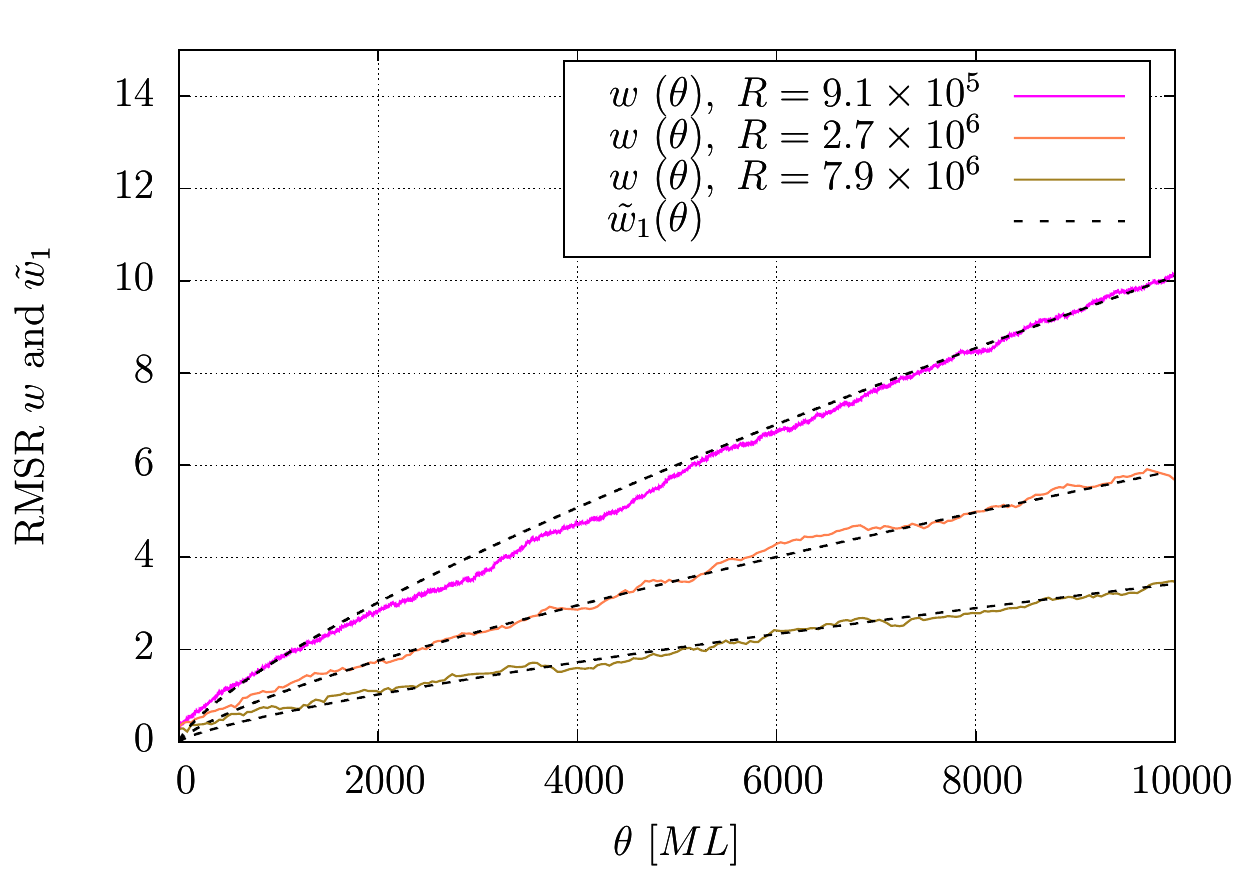} \hfill
\includegraphics[width=0.49\hsize]{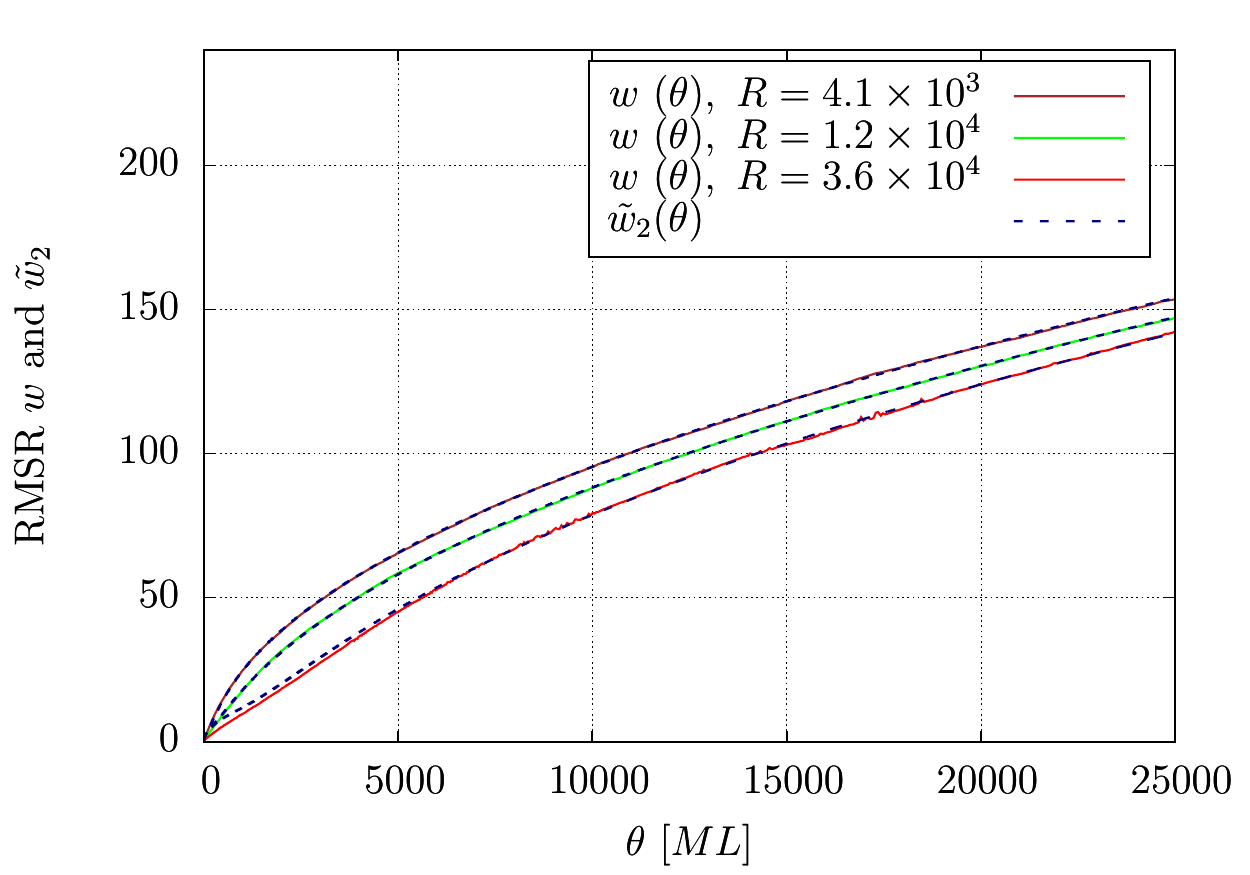}
\caption{\label{fig-both}{Approximation of $w$ by $\tilde{w}_1$ for higher values of $R$ (left) and  by  $\tilde{w}_2$ for smaller $R$ (right).}} 
\end{centering}
\end{figure}

\begin{figure}[ht!]
\begin{centering}
\includegraphics[width=0.4\hsize]{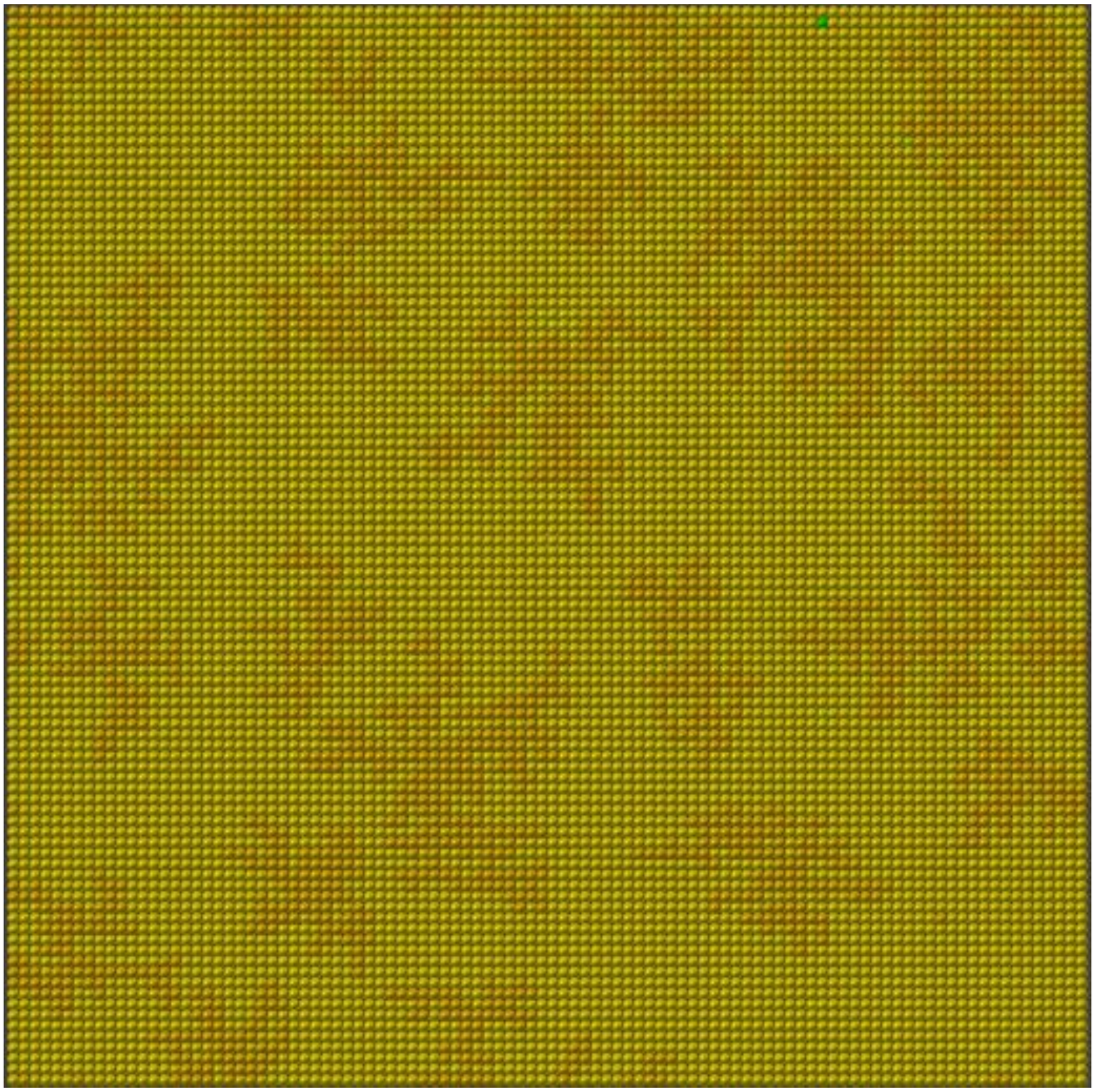} \hfill
\includegraphics[width=0.4\hsize]{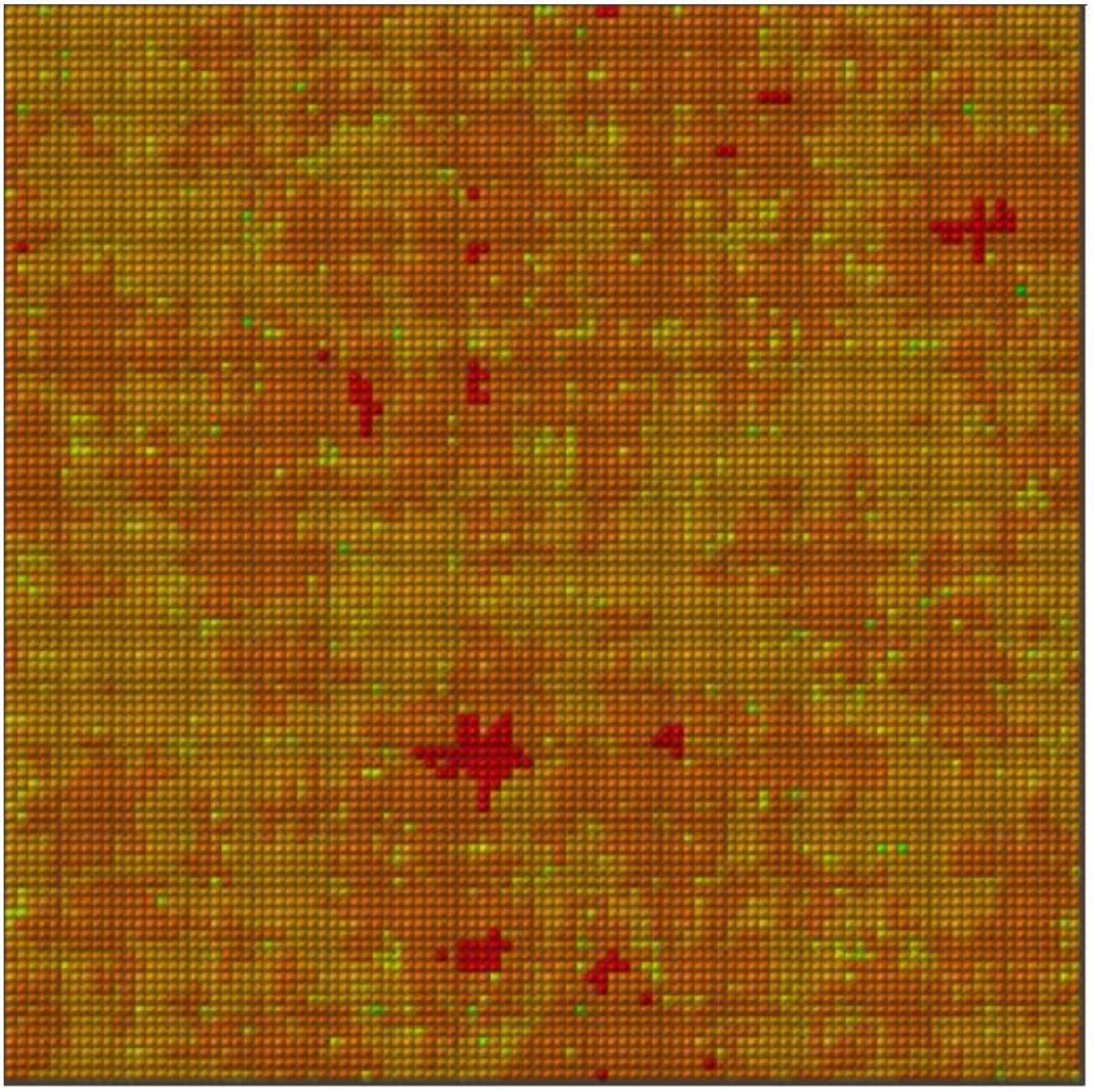}
\caption{\label{fig-snap} Snapshots
of the surface  at coverage $\theta=17$\,ML for $R=\numm{9.1E+05}$ (left) and  $\theta=14$\,ML for $R=\numm{4.1E+03}$ (right). The color changes from green to red with increasing the height.} 
\end{centering}
\end{figure}

 
Least squares approximations provide the following analytic expressions of $w$ in the transient regime with $\beta =0.75$ and in the asymptotic regime with $\beta =0.5$, correspondingly:
\be  
\tilde{w}_1(R,\theta)=9.64R^{-1/2}\theta ^{3/4}
\ee
and
\be  
\tilde{w}_2(R,\theta)=g(\theta;\lambda,\xi)\theta ^{1/2},
\ee
where the function $g$ is defined by 
\be
g(\theta;\lambda,\xi)=\exp[\exp[-(\log \theta/{\lambda})^{\xi}]-1],
\ee 
 the parameters 
$\lambda$ and $\xi$ depend only on $R$ and some of their values 
are presented in Table~\ref{tab1}. However, it should be emphasized that the function $g$ used in the definition of $\tilde{w}_2$ is not unique, the values of $\lambda$ and $\xi$ are approximate and may undergo slight changes if results from longer simulations are used.
\begin{table}[ht!]
\begin{centering}
\begin{tabular}{|c|c|c|}
\hline
$R$ & $\lambda$ & $\xi$ \\ 
\hline\hline
\numm{4.7E+02}& 3.60  & -3.54 \\
\hline
\numm{1.4E+03}& 4.27  & -3.96 \\
\hline
\numm{4.1E+03}& 5.61  & -6.12 \\
\hline 
 \numm{1.2E+04}& 6.69 & -6.29 \\ 
\hline 
 \numm{3.6E+04}& 7.99 & -9.12  \\ 
\hline
 \numm{1.0E+05}& 9.81 & -14.37 \\ 
\hline
\end{tabular} 
\caption{\label{tab1} The values of $\lambda$ and $\xi$ used in the definition of  $g(\theta;\lambda,\xi)$.}
\end{centering}
\end{table}
In Fig.~\ref{fig-tot} (left),  simultaneous approximation of $w$ by 
$\tilde{w}_1$ and $\tilde{w}_2$ is shown in the cases $R=\numm{3.6E+04}$ and $R=\numm{1.0E+05}$ (the only ones in our simulations with $L=100$ where both periods with $\beta =0.75$ and $\beta=0.5$ are clearly visible). 
In the right part of the same figure, only the functions $\tilde{w}_1$ are shown for three different values of $R$.
Fig.~\ref{fig-both} also presents approximations of $w$  by $\tilde{w}_1$ for higher deposition-to-diffusion ratio $R$ (left) and by $\tilde{w}_2$ for lower $R$  (right).
These observations are clear evidences  that the scaling behavior of the epitaxial model under consideration is described in asymptotic limit by
\be
w(R,\theta)\sim \theta ^{1/2},
\ee
that is, for $\theta \rightarrow \infty$ it coincides with the model of random deposition. 
 Furthermore, it possesses a pronounced  (pre-asymptotic) transient regime   obeying the law
\be 
\label{w2-law-2}
{w}(R,\theta)\sim R^{-1/2}\theta ^{3/4}.
\ee
Not surprisingly, the prefactor $R^{-1/2}$  in Eq.~\ref{w2-law-2} is the same as in the analytic formula for $w$ derived from traditional SOS model with Arrhenius dynamics (see \cite[page 4]{aar}), which is an indirect evidence for the validity of our model. 
As we mentioned in the introduction, due to this  prefactor, at low temperatures or for small $R$ the surface is covered by a large number of unevennesses  and vice versa -- extremely low concentration of such irregularities is observed if $R$ is large enough. For comparison,   snapshots of the surface can be seen in Fig.~\ref{fig-snap} at coverages $\theta=17$\,ML for $R=\numm{9.1E+05}$ (left) and  $\theta=14$\,ML for $R=\numm{4.1E+03}$ (right). 
\section{Some comments on growth exponents}
\label{sec4}
 A careful inspection of the surface indicates that after a short initial period of planar growth, it gradually fills with "holes" (or "valleys") due to the imposed infinite ES barrier for steps of height greater than 1. The difference between the two regimes ($\beta=3/4$ and $\beta=1/2$) as well as the crossover point is determined by the fractional part of total area of such hollows from the substrate area. The first regime is marked by formation of (predominantly, narrow) porous pillars where the material supply is possible only by adsorption. With increasing time their fractional part increases, thus, the importance of adsorption events increases and the surface growth process enters the second regime (approaching random deposition) which can be considered also as a second transient regime before the "saturation steady state" which shrinks to 0 for $\alpha=\infty$.
On the other hand, the first period ($\beta=3/4$) is quite unusual and similar examples with scaling exponents larger than $1/2$ have already been observed in the papers \cite{kess,ariz} proposing quite different models of MBE growth. In \cite{kess}, the model of ballistic aggregation with surface diffusion is introduced, where the deposited particle sticks either on  top of a column or at one of its sides.
Then the particle can move to a new position chosen among the best-coordinated sites in a hypercubical box. The simulations with this model 
are characterized with three different regimes preceding the saturation - in the first, representing the surface diffusion scaling, $w$ grows as $t^\beta$, $\beta=0.25$, there follows a period of rapid growth, and the third power-law regime and subsequent saturation are described by KPZ equation. The authors underline the interesting fact that the period with sharp rise in width is governed only by the size of hypercubical box. Similar growth profiles are obtained in \cite{ariz} with another, rough-surface MBE model based on  a second order continuum equation
and a corresponding  difference equation using the discrete local curvature. The difference equation depends on a threshold $Z=rh({\mathbf x},t)$, $r$ being a parameter. As in \cite{kess}, an intermediate period of rapid growth with an exponent dependent on $r$ but larger than 0.5  is observed in simulations. The only difference with \cite{kess} is that like in our model, simulations with this rough-surface model approach random deposition regime as $t\rightarrow\infty$.  
\par
It should also be noted that our model, being a representative of random deposition class, can be considered as a "boundary case" in the family of {\em super-rough} models \cite{sarm2,sarm3} characterized with $\alpha >1$. Though these models do not provide genuine asymptotic pictures of real physical surface growth problems, they may very well describe short-time (non-asymptotic) kinetic surface roughening in many experimental processes \cite{sarm4}.   
For example, the surface profile in the first transient regime is characterized with deep holes ranging from narrow pillars (for large $R$) to wide grooves (for relatively small $R$) and at suitable temperatures seems to be comparable to the surfaces of certain porous materials (e.g. substrates for low-$k$ materials in microelectronics \cite{bakl}). Therefore, this model or its natural generalizations could successfully be  used to mimic the epitaxy of such materials.
\par
Though it cannot be elucidated  by the results from our simulations due to the limited range of $R$, one could expect that  further increasing of $R$ would lead to shifting the pre-asymptotic period with $\beta=3/4$ to the right and give rise to another (preceding it) transient regime corresponding to VLDS class ($\beta=1/5$, see \cite{aar}) or, in other words -- the model would approach the VLDS class as $R\rightarrow \infty$.
\par 
The first results from simulations with included nearest-neighbor interaction energy in the definition of activation barrier $\Delta E$ exhibit the same characteristics as in the case of irreversible growth -- allowing detachment of atoms from islands leads only to prolongation of the pre-asymptotic regime with growth exponent $\beta=3/4$. 
\section{Conclusions}
We have proposed an alternative of the commonly used solid-on-solid model with Arrhenius dynamics.
The results from first simulations with our model show that kinetic roughening of the surface 
for intermediate values of diffusion-to-deposition ratio $R$ is successfully described by two stable growth regimes -- a transient pre-asymptotic regime with growth exponent $\beta =3/4$ and an asymptotic state with $\beta=1/2$ (coinciding with random deposition).
Our considerations and precise results derived from simulations confirm the validity and applicability of the presented KMC model. We have performed also, but not shown in this work, some simulations with the extended definition of $\Delta E$ including nearest-neighbor interactions which seemingly do not alter the profile of temporal roughness evolution.


\begin{thebibliography}{}
\expandafter\ifx\csname url\endcsname\relax
  \def\url#1{\texttt{#1}}\fi
\expandafter\ifx\csname urlprefix\endcsname\relax\def\urlprefix{URL }\fi
\expandafter\ifx\csname href\endcsname\relax
  \def\href#1#2{#2} \def\path#1{#1}\fi

\bibitem{fami1}
F.~Family, T.~Vicsek, J. Phys. A: Math. Gen. {\bf 18}, L75 (1985)

\bibitem{sarm}
S.Das Sarma, P.~I. Tamborenea, Phys. Rev. Lett.  {\bf 66}, 325 (1991)

\bibitem{krug3}
J.~Krug, Phys. Rev. B {\bf 52}, 8550 (1995)

\bibitem{smil2}
P.~{\v{S}}milauer, M.~Kortla, Phys. Rev. B {\bf 49}, 5769 (1994)

\bibitem{bart1}
M.~C. Bartelt, J.~W. Evans, Phys. Rev. Lett {\bf 75}, 4250 (1995)

\bibitem{anna}
A.~Chame, F.~D.~A. Aar{\~a}o, Surf. Sci. {\bf 553}, 145 (2004)

\bibitem{henk}
G.~Henkelman, H.~J{\'{o}}nsson, Phys. Rev. Lett. {\bf 90}, 116101 (2003)

\bibitem{andr}
A.~Patrykiejew, K.~Binder, Surf. Sci. {\bf 273}, 413 (1992)

\bibitem{marm}
I.K. Marmorkos, S. Das Sarma, Phys. Rev. B {\bf 645}, 11262 (1992)

\bibitem{kond}
J.~Kondev, C.~L. Henley, D.~G. Salinas, Phys. Rev. E {\bf 61}, 104 (2000)

\bibitem{mull}
W.W. Mullins, J. Appl. Phys. {\bf 28}, 333 (1957) 

\bibitem{herr}
C.~Herring, in: W.~E. Kingston (Ed.), The Physics of Powder Metallurgy,
  McGraw-Hill, New York, 1951.

\bibitem{edwa}
S.F. Edwards, D.~R. Wilkinson, Proc. R. Soc. London, Ser. A {\bf 381}, 17 (1982)

\bibitem{kardar1}
M.~Kardar, G.~Parisi, Y.-C. Zhang, Phys. Rev. Lett. {\bf 56}, 889 (1986) 

\bibitem{vill1}
J.~Villain, J. Phys. I {\bf 1}, 19 (1991) 

\bibitem{lai}
Z.-W. Lai, S. Das Sarma, Phys. Rev. Lett. {\bf 66}, 2348 (1991) 

\bibitem{weeks1}
J.D. Weeks, G.H. Gilmer, K.A. Jackson, J. Chem. Phys. {\bf 65}, 712 (1976) 

\bibitem{evans1}
J.W. Evans, Phys. Rev. B {\bf 39}, 5655 (1989)

\bibitem{fami2}
F.~Family, J. Phys. A {\bf 19}, L441 (1986)

\bibitem{brune}
H.~Brune, Surf. Sci. Rep. {\bf 31}, 121 (1998) 

\bibitem{mich}
T.~Michely, J.~Krug, Islands, Mounds, and Atoms, Springer, Berlin, 2004.

\bibitem{evans2}
J.W. Evans, P.A. Thiel, M.C. Bartelt, Surface Science Reports {\bf 61}, 1 (2006)
 
\bibitem{evans3}
H.~C. Kang, J.~W. Evans, Surf. Sci. {\bf 271}, 321 (1992)

\bibitem{kardar2}
M.~Kardar, Pysica A {\bf 281}, 295 (2000) 

\bibitem{tamb}
P.I. Tamborenea, S. Das Sarma, Phys. Rev. E {\bf 48}, 2575 (1993)

\bibitem{krug2}
J.~Krug, M.~Plischke, M.~Siegert, Phys. Rev. Lett. {\bf 70}, 3271 (1993) 

\bibitem{kort}
M.~Kortla, P.~{\v{S}}milauer, Phys. Rev. B {\bf 53}, 13777 (1996) 

\bibitem{wil}
M.R. Wilby, D.D. Vvedensky, A.~Zangwill, Phys. Rev. B {\bf 46}, 12896R (1992)

\bibitem{has1}
C.A. Haselwandter, D.D. Vvedensky, Phys. Rev. E {\bf 77}, 061129 (2007)

\bibitem{has2}
C.A. Haselwandter, D.D. Vvedensky, Europhys. Lett. {\bf 77}, 38004 (2008) 

\bibitem{has3}
C.A. Haselwandter, D.D. Vvedensky, Int. J. Mod. Phys. B {\bf 22}, 3721 (2008) 

\bibitem{aar}
F.D.A. Aar{\~a}o, Phys. Rev. E {\bf 81}, 041605 (2010) 

\bibitem{kim1}
J.M. Kim, J.M. Kosterlitz, Phys. Rev. Lett. {\bf 62}, 2289 (1989) 

\bibitem{kim2}
Y.~Kim, D.K. Park, J.M. Kim, J. Phys. A: Math. Gen. {\bf 27}, L533 (1994) 

\bibitem{reif}
F.~Reif, Statistical and thermal physics, McGraw-Hill, New York, 1965

\bibitem{kess}
D.A. Kessler, H. Levine, L.M.Sander, Phys. Rev. Lett. {\bf 69}, 100 (1991)

\bibitem{ariz}
C.M. Arizmendi, J.M. Sanchez, J. Phys.: Condens. Matter {\bf 5}, A103 (1993)

\bibitem{sarm2}
S. Das Sarma, S.V. Ghaisas, J.M. Kim, Phys. Rev. E {\bf 49}, 122 (1994)

\bibitem{sarm3}
S. Das Sarma, C.J. Lanczycki,  R. Kotlyar, S.V. Ghaisas, Phys. Rev. E {\bf 53}, 359 (1996)

\bibitem{sarm4}
S. Das Sarma, in: Z. Zhang and M.G. Lagally (Ed.), Morphological Organization in Epitaxial Growth and Removal, World Scientific, Singapore, 1998. 

\bibitem{bakl}
M.R. Baklanov, K. Maex, Phil. Trans. R. Soc. A {\bf 364}, 201 (2006)

\end{thebibliography}
\end{document}